# Title: Self-organization of stress patterns drives state transitions in actin cortices

**Authors:** Tzer Han Tan[1+], Maya Malik Garbi[2+], Enas Abu-Shah[2,3+#], Junang Li[1], Abhinav Sharma[4,6], Fred C. MacKintosh[4], Kinneret Keren[2,3,5*], Christoph F. Schmidt[6,7*], Nikta Fakhri[1,6*]

[*] Corresponding authors

[+] These authors contributed equally

[#] Current address: Kennedy Institute of Rheumatology, University of Oxford, Oxford, UK

**Affiliations:**

[1] Department of Physics, Massachusetts Institute of Technology, Cambridge, MA, USA

[2] Department of Physics, Technion – Israel Institute of Technology, Haifa, Israel

[3] Russell Berrie Nanotechnology Institute, Technion – Israel Institute of Technology, Haifa, Israel

[4] Department of Physics and Astronomy, Vrije Universiteit, Amsterdam, The Netherlands

[5] Network Biology Research Laboratories, Technion – Israel Institute of Technology, Haifa, Israel

[6] Third Institute of Physics – Biophysics, Georg August University, Göttingen, Germany

[7] German Center for Cardiovascular Research (DZHK), Göttingen, Germany

**Abstract**

**Biological functions rely on ordered structures and intricately controlled collective dynamics. In contrast to systems in thermodynamic equilibrium, order is typically established and sustained in stationary states by continuous dissipation of energy. Non-equilibrium dynamics is a necessary condition to make the systems highly susceptible to signals that cause transitions between different states. How cellular processes self-organize under this general principle is not fully understood. Here, we find that model actomyosin cortices, in the presence of rapid turnover, display distinct steady states, each distinguished by characteristic order and dynamics as a function of network connectivity. The different states arise from a subtle interaction between mechanical percolation of the actin network and myosin-generated stresses. Remarkably, myosin motors generate actin architectures, which in turn, force the emergence of ordered stress patterns. Reminiscent of second order phase transitions, the emergence of order is accompanied by a critical regime characterized by strongly enhanced strain fluctuations. The striking dynamics in the critical regime were revealed using fluorescent single-walled carbon nanotubes as novel probes of cortical dynamics.**

**One Sentence Summary**

Actin cortices display transitions between steady states with distinct structures and stress pattern symmetries.

## Main Text

Phases of matter are typically characterized by symmetry or collective order of its components. Changes in this order or symmetry often correspond to phase transitions in classical equilibrium statistical physics (*1*). Dynamic functions of living systems also require specific forms and collective order. However, in contrast to equilibrium systems, organisms build ordered structures through the dissipation of metabolic energy (*2-4*). A generalization of equilibrium concepts can provide insight into how order emerges in these non-equilibrium systems.

Dissipative cellular structures have to be robust and form stable steady states lasting from seconds to the lifetime of the organism. At the same time, they have to be highly adaptive and reorganize in response to internal or external signals. A classical case of collective dynamics with high sensitivity to control parameters is observed in systems near a critical point or second-order phase transition. Such transitions are characterized by the continuous development of order, which distinguishes two phases (*1*). At such a transition, the susceptibility to certain control parameters diverges, and the system displays large fluctuations with extended spatial and temporal correlations. Criticality is emerging as an important functional feature in biological systems, with regulatory mechanisms positioning systems near criticality (*5*).

A prominent example of an adaptive dissipative structure is the cell cortex. The cortex is a quasi-2D network of dynamically crosslinked actin filaments that continuously polymerize and depolymerize (*6*). This micron thick network is anchored to the cell membrane and is internally activated by myosin motor proteins. The cortex provides mechanical integrity and rigidity to cells over extended timescales, while its components turn over within minutes (*7, 8*). In addition, the cortex undergoes dramatic reorganizations during cellular processes such as division, generation of polarity and motility (*9-11*). These processes often entail transitions from homogeneous to inhomogeneous states via network contraction (*12, 13*). How does the cortex achieve the seemingly impossible feat of switching between stability and large-scale reorganization?

Cortex structure and dynamics are controlled by a subtle interplay between actin turnover, network connectivity, as well as strains and stresses generated by non-muscle myosin motors (*14-16*). Characteristic time scales are on the order of seconds for all these processes (*14, 17*). Experimental access to the active mechanics of cortices has been severely limited. Techniques such as AFM indentation (*18*) or pipette aspiration (*19*) only provide global, coarse-grained measurements, not capturing spatial inhomogeneity, anisotropy and time-dependence (*14*). Microrheology (*19*) could, in principle, provide local information, but its use in the cortex has been limited by the lack of suitable probes; micron-sized colloidal beads are too large to penetrate the thin cortex without perturbing it, while nanometer-sized particles, such as quantum dots, are too small to remain in the network for long enough to probe it.

Here, we use fluorescent single-walled carbon nanotubes (SWNTs) as probes to demonstrate that a subtle interaction between mechanical percolation of the network and propagation of myosin-generated stresses leads to structurally and dynamically distinct steady states in model actin cortices. Transitions between these states are accompanied by striking changes in the symmetry of stress patterns, including a critical regime characterized by strongly enhanced strain fluctuations.

We formed dynamic cortices, *in vitro,* by encapsulating *Xenopus* egg extract in water-in-oil emulsion droplets (Fig. 1) (*20*). Localization of ActA protein to the water-oil interface activated Arp2/3-mediated nucleation of branched actin networks (Fig. 1B), generating homogeneous, ~ 1 µm thick, cortical actin networks. These networks exhibited continuous actin turnover with a typical time scale of ~ 1 min (Fig. S1) as well as myosin-driven network dynamics, in a stable cell-like geometry. Droplets were flattened and confined between hydrophobic coverslips for observation. The droplets were ~ 100 µm in diameter and had a fixed height of 30 µm. The cortical actin networks were imaged by confocal microscopy (Fig. 1A). To resolve network dynamics at higher resolutions, we used near-IR fluorescent SWNTs as novel “stealth” probes (Fig. 1B). SWNTs are ~ 1 nm in diameter and ~ 100 - 300 nm in length (*17*). We found that SWNTs easily penetrate the thin cortex without perturbing the network and, due to their large aspect ratio (*21*), get entrapped for long enough times to report on network dynamics. Due to their extreme photostability, the SWNTs could be tracked over a broad window of time scales (milliseconds to hours).

In droplets of bare extract, we observed homogeneous cortices (Fig. 1A) with no discernable network movement. Since myosins can only drive large-scale movement when the network is sufficiently crosslinked, we increased network connectivity by adding an actin crosslinker, α-actinin. Increasing crosslinker concentration in this system leads to breaking of symmetry and formation of a polar cap (*20*). We found that, as a function of crosslinker concentration, we could drive the system into three dramatically different dynamic steady states, referred to as low, intermediate and high connectivity (Fig. 2). In all states, continuous actin turnover was maintained (Fig S1).

At low connectivity (0 - 1.5 μM added α-actinin), cortices remained homogeneous, and SWNTs fluctuated randomly (Fig. 2A, Movie S1). The homogenous actin distribution reflects a dynamic steady state, with continuous actin polymerization catalyzed by the ActA at the surface and depolymerization to the bulk (Fig. S1). A control experiment with myosin immuno-depleted extracts (Fig. S2) showed that the spatially uncorrelated and random probe motions were largely motor driven.

At intermediate connectivity (2.0 - 2.5 μM added α-actinin), coherently moving clusters of SWNTs appeared with diameters up to 10 μm. Strikingly, the clusters moved over long distances within the cortex, without generating large-scale actin density inhomogeneities (Fig. 2B, Movie S2). The clusters moved in a vortical manner, with high directional persistence, at typical speeds of ~ 10 μm/min. This motion was likely driven by contractile forces acting between clusters. Occasional rapid changes of cluster velocities (Movie S3) can be explained by the rupture of tenuous bridges, contracted by few myosin minifilaments. Note that the mechanism of force generation is different from that of active swimmers (*22*), such as fish in a school, where particles propel themselves by exerting a force against an embedding medium or a surface. The dynamics in our model cortices are also different from the observed phenomena in highly concentrated motor activated solutions of actin filaments (*23*) and microtubules (*24*), where order is imposed through steric repulsion between long filaments.

At high connectivity (3.0 - 4.0 μM added α-actinin), cortices phase separated and formed a large cap, which extended over a substantial fraction of the droplet surface, but remained essentially 2-dimensional. Small clusters continuously nucleated and flowed radially towards the cap. This contraction was visible in both the actin and the SWNT fluorescence channels (Fig. 2C, Movie S4 and S5). The velocity of the converging clusters was ~ 10 μm/min in the periphery and decreased towards the cap (Movie S5). The caps formed at random positions, but once formed, remained stable for at least 2 hours. The ratio of intensities between the cap region and peripheral accretion zones also remained constant over time (Fig. S3). This shows that the system is at steady state, maintained by the continuous disassembly of actin from the cap.

A striking feature of these non-equilibrium steady states is that they all display distinct patterns of motion while only the high connectivity state exhibits an inhomogeneous density pattern. Two-dimensional correlation analysis of SWNT velocities shows a rapid but continuous growth of correlation length at intermediate connectivity (Fig. S4). Increasing correlation length reflects increasing connectivity in the actin cortex, and the shape of the curve (Fig. S4) indicates cooperativity and is reminiscent of connectivity percolation (*25*). Breaking up of the network into individual clusters is commonly observed as a consequence of crosslinking, especially in combination with contractile motors (*26, 27*). In contrast to previous systems, in our dissipative steady-state system, the clusters maintain their size but are dynamic, with their lifetime exceeding that of their components, i.e. they display collective structural memory. The most prominent feature we observe is the long distance movement of the clusters driven by myosin motors, demonstrating, in addition, a collective motion memory.

At intermediate connectivity, patterns of motions fluctuate strongly (Movie S2 and Fig. 2B), analogous to fluctuations observed near a critical point of an equilibrium system. To quantify fluctuations, we extract changes of velocity in fixed positions in the cortex. We calculated the normalized coarse-grained velocity fluctuation autocorrelation in time (Fig. 3A), $\tilde{C}_{\Delta}(\tau)=\left\langle \Delta\hat{V}_{\alpha}(t)\cdot\Delta\hat{V}_{\alpha}(t+\tau)\right\rangle_{\alpha T}$. Velocity fluctuation, $\Delta\vec{V}$, is the difference between the velocity in a given coarse-grained box $\vec{V}_{\alpha}(t)$ and the time-averaged velocity in the same box $\left\langle \vec{V}_{\alpha}\right\rangle_{T}$ (See SI). Figure 3A shows the velocity fluctuation autocorrelation in time for a 5x5 grid at the different crosslinker concentrations. Local average velocities are close to zero at low and intermediate connectivity but non-zero and converging to the cap at high connectivity. Fluctuations have small amplitudes and are rapid at low connectivity, but gain amplitude and persistence at intermediate connectivity, to become of small amplitude again at high connectivity. We constructed a scalar order parameter to quantify the persistence of local velocities. The order parameter was calculated as the angular correlation of velocities in the coarse-grained boxes, $\langle\cos\phi\rangle_{50s}$, averaged over time intervals of 50 s, longer than the correlation time of a myosin mini-filament (~ 5 s (*17*)) (Fig. 3B). The order parameter shows a steep increase at intermediate connectivity, analogous to a second order equilibrium phase transition. The increasing order in probe particle velocities reflects spatial ordering of the local stresses, rather than liquid-crystalline nematic structure of the filaments in the cortex. From the relative numbers of homogeneous and phase-separated cortices at a given crosslinker concentration, inferred from images of rhodamine-labeled actin, we constructed an actin-density based structural order parameter (Figs. 3C and S5). This order parameter confirms a global phase separation at high connectivity.

Near critical points, systems exhibit extended correlations over length-scales that substantially exceed molecular scales. To extract the average spatial extent of correlated velocity fluctuations driven by the fluctuating arrangements of stress generating motors, we used the deviations from the overall local velocity averages. The normalized velocity fluctuation correlation function in space (see SI) (Fig. 3D) was calculated as $\tilde{C}_{\Delta}(R)=\left\langle \Delta\hat{v}_{i}(\vec{r}_{i},t)\cdot\Delta\hat{v}_{j}(\vec{r}_{j},t)\delta\left(\left|\vec{r}_{i}-\vec{r}_{j}\right|-R\right)\right\rangle_{ijT}$. Velocity fluctuation, $\Delta\vec{v}_{i}$, is the difference between the velocity of an individual particle $\vec{v}_{i}$ and the overall average velocity in the coarse-grained box $\alpha$ containing particle $i$, $\langle \vec{V}_\alpha \rangle_T$ (See SI). At low connectivity, spatial correlations of velocity fluctuations are small. At high connectivity, deviations from the local average velocity, which is zero in the cap region and non-zero in the periphery, are also small, so that correlations become buried in the noise. At intermediate crosslink densities or marginal connectivity, however, $\tilde{C}_\Delta(R)$ shows a characteristic fluctuation correlation length of ~ 15 µm smaller, than but approaching the system size (Fig. 3E).

Another hallmark of equilibrium critical points is diverging susceptibility, i.e. a diverging response to external driving such as an applied force field (*1*). In the non-equilibrium steady states of the cortices, driving is not external, but caused by myosin motors creating an internal stress field. We define susceptibility as the amount of changes in patterns of motion in response to small changes in the stress field, i.e. the addition or removal of a few myosin motors. Analogous to thermal fluctuations in equilibrium, in our model system, myosin on/off kinetics provides stochastic force fluctuations. Thus we can infer susceptibility from the strain fluctuations driven by these force fluctuations. To obtain a measure of susceptibility from the observed patterns of motion, we calculated the variance of velocity fluctuation correlations at a fixed distance larger than the typical length of a single actin filament (> 1 µm). Figure 3F shows that this variance exhibits a broad peak at intermediate connectivity. Interestingly, a numerical simulation of an isotropically stretched 2D network of linear springs also shows correlated strain fluctuations, i.e. moving clusters, near connectivity percolation (Fig. S6, see SI).

The three states differ in the symmetry of the patterns of motion of the probe particles. To quantify average symmetry and order of the respective states, we calculated directional correlation functions for the entire sets of data (Fig. 4). This function was constructed by calculating the velocity-velocity directional correlation for each moving probe, with all the surrounding probes, as a function of distance and angle with respect to its own direction of motion, and then averaging over all moving particles in all movies (*28*) (see SI).

At low connectivity, patterns are fully symmetric, and no correlation is observed (Figs. 4A and S7). At intermediate connectivity, the vector field of probe velocities begins to show curl but no divergence (Figs. 4B and S7). The inner region of positive correlation in Fig. 4B reports average cluster size, consistent with cluster size determined from displacement correlations (Fig. S8). The breaking of rotational symmetry is evident from the extended positive correlations around 0° and 180°, i.e. a finite probability that particles trailing and leading a given particle move in the same direction. Such a breaking of rotational symmetry is characteristic for nematic liquid crystals. Note that in this case, ordering of velocities implies a polar nematic (*22, 29*). In our system, order is not generated by alignment of actin filaments through entropy maximization, as is typically the case for equilibrium nematics. Rather, the aligned velocity vectors likely reflect underlying ordering of contractile myosin minifilaments that can be conceptualized as force dipoles. At high connectivity, the velocity vector field around the cap shows divergence, rather than curl, i.e. it reflects broken translational symmetry (Figs. 4C and S7). The directional correlation function clearly shows first, an extended correlation length or cap size, and second, an asymmetry between 0° and 180° which is a signature of a converging flow field (Fig. S9). This converging velocity field likely reflects a radial stress pattern, i.e. a radial orientation of myosin force dipoles. Such global centrosymmetric polar order is not observed in typical liquid crystals. The crucial difference in our system is the lack of conservation of particles because of continuous actin depolymerization in the cap and recycling of monomers to the bulk.

Figure 4 (D-F) presents a conceptual model for the observed non-equilibrium steady states. The generation of order and the breaking of symmetry in the cortex are driven by the interplay of network connectivity and active stresses. Increasing connectivity drives the network towards mechanical percolation, which acts like a clutch allowing myosin motors to exert increasingly long-range forces. This leads to long-range transport in the cortex and structural rearrangements, which in turn, feeds back on the arrangement of the force-generating myosin motors. The observed nematic and centrosymmetric polar orders in probe velocities thus reflect the order in the orientation of the stress generators (myosin minifilaments). At low connectivity (Fig. 4D), myosin will engage actin, but will not drive correlated motions over distances longer than filament length. At intermediate connectivity (Fig. 4E), the effective range of stress propagation expands. The combination of contractility and crosslinking creates relatively rigid islands of limited size, where the density is not different enough to be visible by actin fluorescence, but the mobility of probe particles within clusters is strongly suppressed. Myosin-activated contractile bridges move the clusters in random and changing directions. Note that competing motors can also accelerate bridge rupture, creating a negative feedback mechanism for cluster growth (Movie S3). We speculate that in our case, in the presence of constant turnover, cluster size ($r_c$) is primarily limited by the competition between accretion at the boundary ($\propto r_c$) and depolymerization within the cluster ($\propto r_c^2$). At high connectivity (Fig. 4F), a transition occurs when the largest cluster reaches system sizes and concentrates actin filaments, motors, and crosslinkers, such that a radial gradient is likely to emerge. This creates net polymerization in the periphery and net depolymerization in the cap, sets up a steady flow of small clusters towards the cap, and in turn orients the motor dipoles radially in the periphery. This model predicts a system-size dependence of the second transition.

The observed dynamics highlight some of the unique properties of “active matter”: active stresses and mechanical structures can couple in complex and subtle ways such that drastically different dynamic patterns can result from shifts in the balance between competing molecular time scales. The intermediate case of marginal connectivity resembles a critical state, albeit in a non-equilibrium situation, and shows maximal susceptibility to internal stress variations. This critical state extends over a relatively broad crosslinker concentration range, which is a feature of a robust state. This robustness is a hallmark of self-organized criticality (*26, 30*). In vivo dynamic transitions such as cortical symmetry breaking in *C. elegans* (*9, 10*), the transition into a contractile ring in oocytes (*31*), membrane stirring (*32*) or the formation of immunological synapses (*33*) bear striking resemblance to our observations. This suggests that the same non-equilibrium mechanisms are at work in vivo. Both the connectivity of the network and the activity of myosin motors can be actively tuned by cells in response to various signals such as $Ca^{2+}$ ions (*15, 34*). With SWNTs as ideal probes, it now remains to tackle the much more challenging problem of mapping cortical dynamics in living cells.

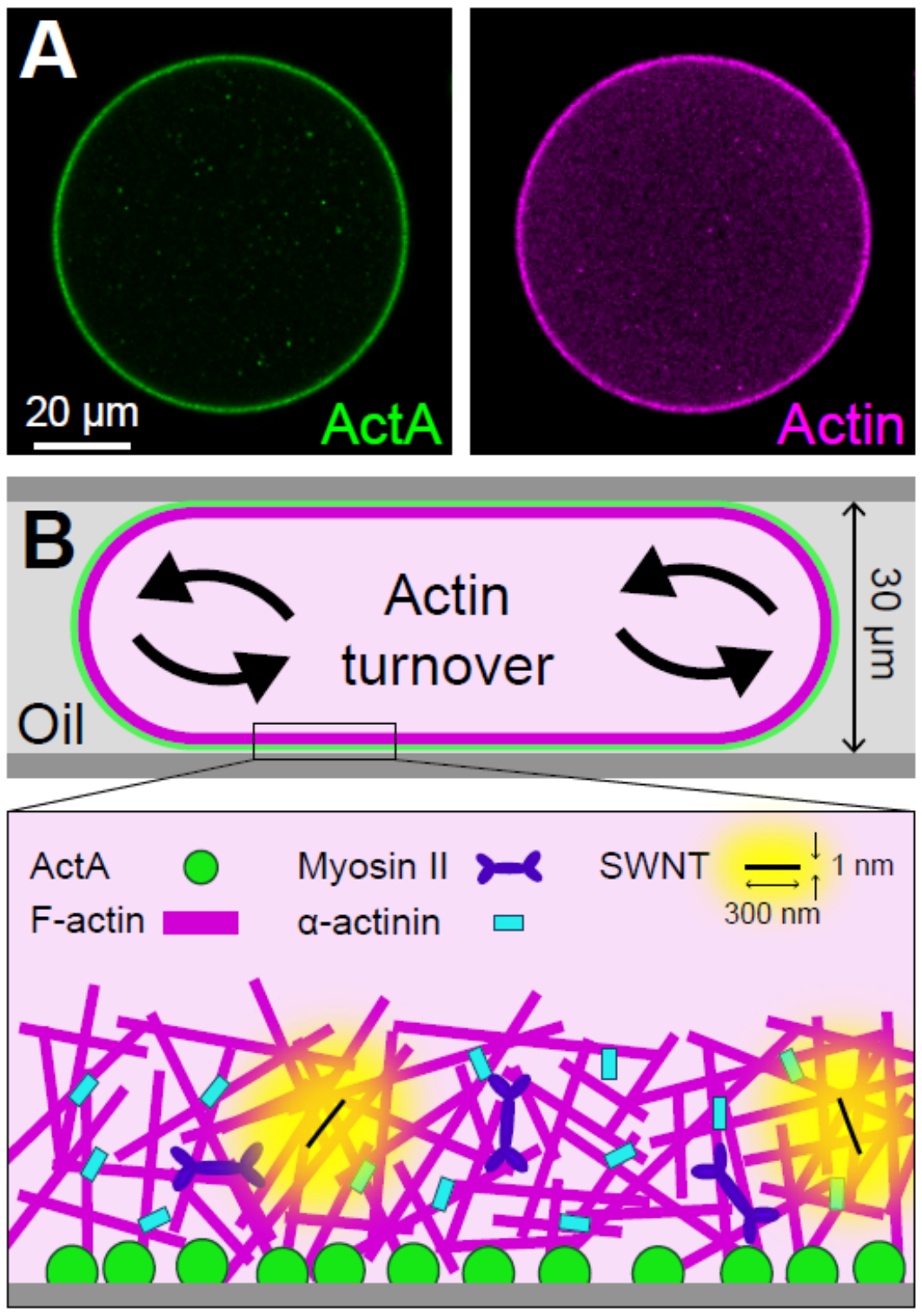

**Figure 1. Dynamic model actin cortices**
Quasi-2D actin networks were generated by encapsulating *Xenopus* egg extract in water-in-oil emulsion droplets and confining actin polymerization to the interface. (A) Equatorial cross section of flattened droplets. Simultaneous confocal imaging of bodipy-conjugated ActA (green, left) and rhodamine-labeled actin (magenta, right). Amphiphilic bodipy-ActA localizes to the water-oil interface and catalyzes the formation of a quasi 2D dynamic actin network by local activation of Arp2/3. (B) Top: Schematic of the experiment. Cortical dynamics are tracked near the flat bottom surface of the droplet. There is continuous turnover between the thin polymeric actin layer and actin monomers in the bulk (arrows). Bottom: Zoomed in schematic of the essential components of the cortex. IR fluorescent SWNTs are inserted as probes of cortex dynamics.

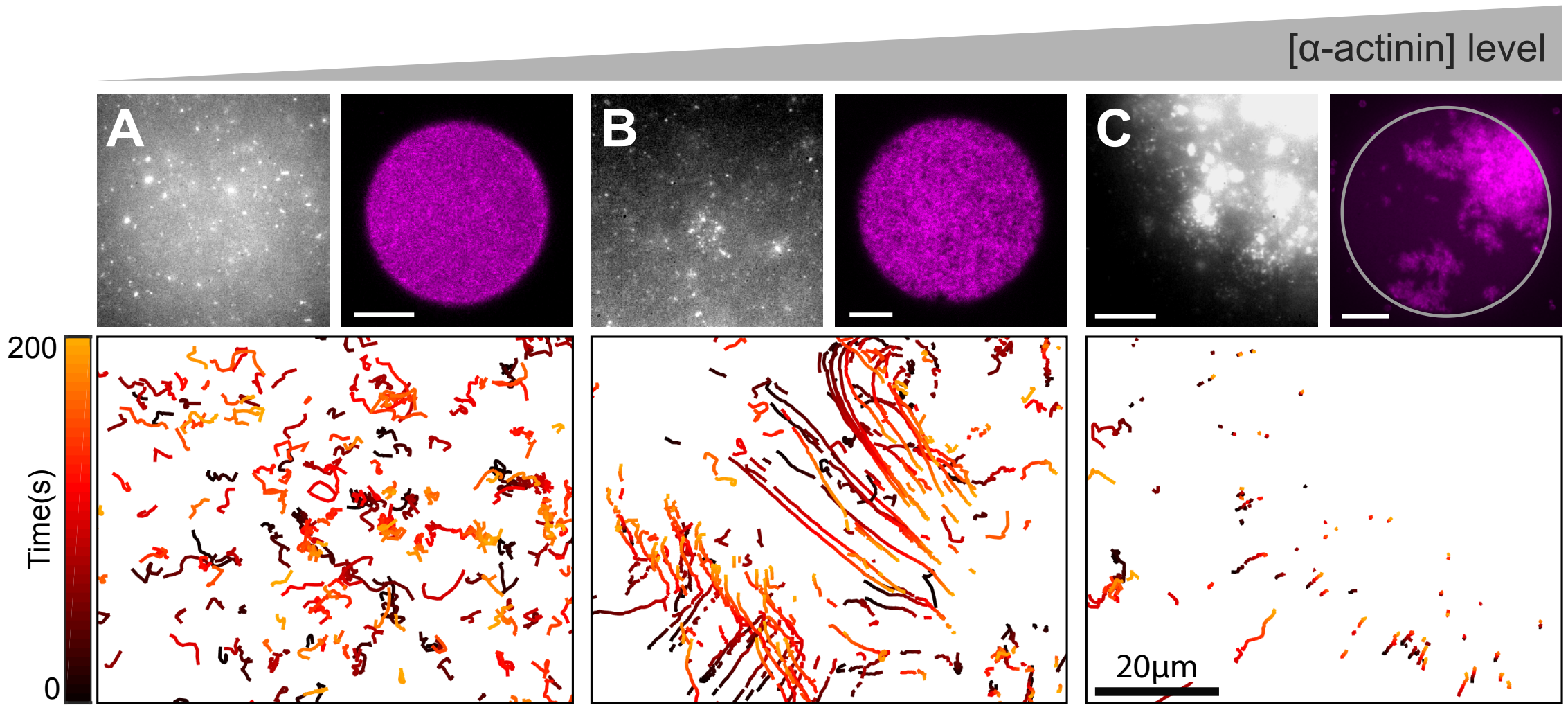

**Figure 2. Signatures of distinct steady states.**
Network percolation can be induced by increasing crosslinkers density (α-actinin). Three distinct steady states can be identified based on qualitative differences in collective network dynamics imaged in the bottom cortical layer in the droplets. (A) Low crosslinking (0-1.5 μM added α-actinin, images shown: 1.5 μM) (B) Intermediate crosslinking (2-2.5 μM added α-actinin, images shown: 2.5 μM). (C) High crosslinking (3-4 μM added α-actinin, images shown: 3 μM). (A-C) Top left: Fluorescence image of inserted SWNT probes. Top right: Confocal image of the actin distribution. Bottom: Individual SWNTs were tracked in movies of 200 s long with 2000 frames (Movies S1-S4). Tracks of individual SWNTs color-coded for progression in time. (Scale bar = 20 μm)

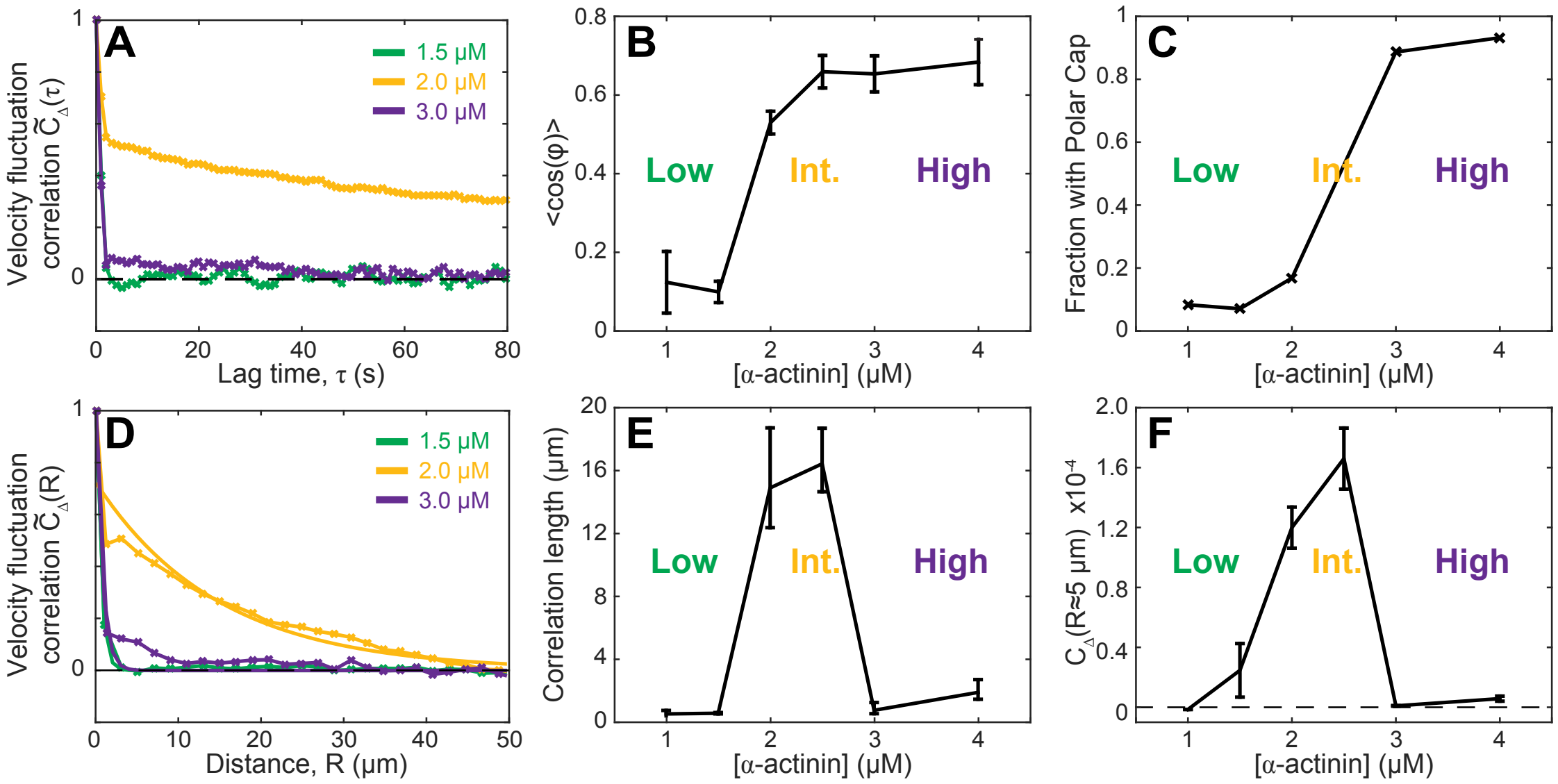

**Figure 3. Collective dynamics and critical fluctuations.**
Steady states differ in the degree of order and correlation of probe particle velocities. (A) Normalized velocity fluctuation autocorrelation is plotted as a function of lag time for different α-actinin concentrations evaluated in a 5x5 coarse grained grid for 200 s movies, averaged over all grid points. (B) Order parameter calculated from the averaged velocity directional persistence as a function of α-actinin concentration. (C) An actin density-based structural order parameter showing the fraction of phase-separated cortices as a function of crosslinker concentration (0-4 μM of added α-actinin), inferred from images of actin fluorescence. (D) The normalized velocity-velocity fluctuation cross-correlation at zero lag time is plotted as a function of probe distances for different α-actinin concentrations (symbols). Data are fitted with single exponentials (lines). (E) Characteristic correlation lengths from fits in (D) as a function of α-actinin concentrations. Error bars: 95% confidence interval. (F) Susceptibility of cluster motion to stress fluctuations quantified from the variance of the amplitude of the velocity-velocity fluctuation correlation taken at $R$ = 5 μm, plotted as a function of α-actinin concentration. Error bars correspond to standard deviations.

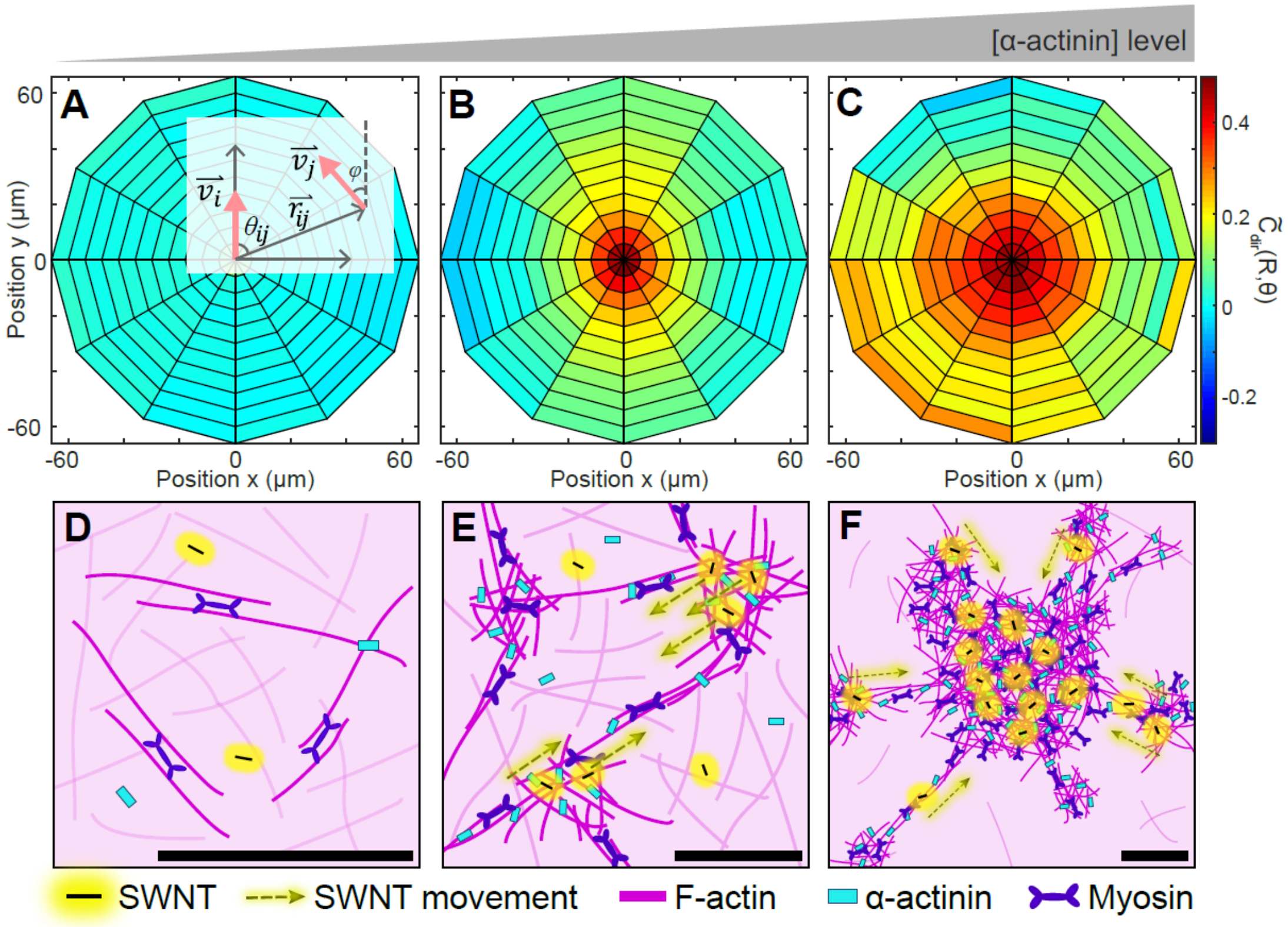

**Figure 4. Order and symmetries in the stress patterns.**
Increasing crosslinkers density leads to increasing order and breaking of symmetries in the arrangement of stress generators. (A-C) Directional velocity-velocity correlation maps the degree of order and shows breaking of symmetry for increasing α-actinin concentrations. For each concentration, the colors portray the average alignment of pairs of probe velocities, as a function of the distance and angle between them (A inset) At low crosslinking (1 μM; A) there is no appreciable correlation. At intermediate connectivity (2.5 μM; B), the map shows enhanced correlations, distributed anisotropically, primarily along the front-back axis, signifying the emergence of nematic polar order, and the breaking of rotational symmetry. At high connectivity (3 μM; C), the converging network flow leads to an asymmetry between top and bottom, indicating the appearance of centrosymmetric polar order and the breaking of translational symmetry. (D-F) Schematic illustration of the conceptual model for cortex dynamics as a function of connectivity.  At low connectivity (D), the network is not mechanically connected, so myosin cannot generate long-range forces, and the network fluctuates randomly. At intermediate connectivity (E), the network reaches the mechanical percolation threshold. Myosin-generated stresses can propagate over larger distances, leading to the formation of rigid clusters, which move in a vortical manner. The interplay between stresses and network rearrangement lead to local nematic ordering of the force-generating myosin minifilaments. At high connectivity (F), the network self organizes into a single large cap with persistent converging flow of clusters into the cap, and polar ordering of the myosin force dipoles.

**Acknowledgements**

We thank Alexander Solon, Jacques Prost, Sriram Ramaswamy, Mehran Kardar and Alex Mogilner for discussion. This research was supported by a Human Frontier Science Program Fellowship (N.F.), J.H and E.V. Wade Fund Award (N.F.), the Cluster of Excellence and DFG Research Center Nanoscale Microscopy and Molecular Physiology of the Brain (CNMPB) (C.F.S.), the European Research Council Advanced Grant PF7 ERC-2013-AdG, Project 340528 (C.F.S), the Deutsche Forschungsgemeinschaft (DFG) Collaborative Research Center SFB 937 (Project A2) (C.F.S.) and a grant from the Israel Science Foundation (K.K.).